\documentclass[useAMS,usenatbib]{mn2e}
\usepackage{graphicx}

\def\la{\raise.5ex\hbox{$<$}\kern-.8em\lower 1mm\hbox{$\sim$}}
\def\ma{\raise.5ex\hbox{$>$}\kern-.8em\lower 1mm\hbox{$\sim$}}

\def\msol{M$_{\odot}$ }
\def\Lsol{L$_{\odot}$ }
\def\kms{$\rm km\, s^{-1}$}
\def\cm3{$\rm cm^{-3}$}
\def\Ts{$\rm T_{*}$~}
\def\Vs{$\rm V_{s}$~}
\def\n0{$\rm n_{0}$}
\def\B0{$\rm B_{0}$}
\def\ne{$\rm n_{e}$~}

\def\Te{$\rm T_{e}$}

\def\erg{$\rm erg\, cm^{-2}\, s^{-1}$}
\def\mum{$\mu$m~}

\def\L12{L$_{12\mu m}$~}
\def\F12{F$_{12\mu m}$~}

\def\Hb{H${\beta}$~}
\def\Ha{H$\alpha$~}
\def\Ly{Ly$\alpha$~}

\title[The NGC 3393 merger]{Evidences of  merging in the Seyfert galaxy NGC 3393  revealed by modelling the spectra} 
\author[M. Contini]{ M. Contini 
\\
School of Physics and Astronomy, Tel Aviv University, Tel Aviv
69978, Israel \\
}

\begin{document}

\date{Accepted: Received ; in original form 2010 month day}

\pagerange{\pageref{firstpage}--\pageref{lastpage}} \pubyear{2009}

\maketitle

\label{firstpage}

\begin{abstract}
The discovery of two active black holes in the Seyfert galaxy NGC 3393,
separated by  about 490 light years, revealed a merging event.
 This led us to look for other evidences
of galaxy collision and merging through the analysis of the observed  spectra in  different
frequency ranges. We found  preshock densities  higher by a factor of  about 10
in  the NGC 3393 NLR than in other AGN and 
patches of ionized matter beyond the  observed NLR bulk. 
They can be  explained by compression and heating of the gas downstream of shock waves created
by collision. 
Metallicity in terms of the O/H relative abundance, is  about 0.78 solar.
Mg/H depletion by a factor of about 3 compared with solar cannot be explained
by Mg trapping into dust grains, due to  rather high shock velocities.
The low O/H and Mg/H abundances indicate mixing with  external matter during collision.
Twice solar N/H  is predicted by modelling the spectra of  high shock velocity
clouds reached by  a  \Ts$\leq$ 10$^5$ K black-body flux. This suggests that Wolf-Rayet stars could  be created  
by  galaxy collision in the central region.

\end{abstract}


\section{Introduction}

The  new galaxy which  results from  collision of galaxies or from  collision of a galaxy with the ISM 
keeps a few records of the parent objects.  The most  obvious evidence  is the  multiple (generally double) 
AGN when at least two objects are involved. Yet, also the morphological and physical pictures
 across the product galaxy
 show collisional processes involving gas and dust clouds,
  from supersonic flows throughout  the  galactic medium,  up to star birth and 
powerful starbursts, depending on the  characteristics of the colliding parents. 
Theoretical simulation and modelling 
(e.g. Cox et al 2008 and references therein, Hammer et al, 2010) were confirmed by the observations of some mergers
such as  Arp 220, NGC 6240, NGC 7212, etc.

In Seyfert galaxies the ranges of pre- and post-shock densities  and of shock velocities may  show some traces of   collision 
 in some regions inside and outside the narrow line region (NLR) cone 
(e.g.  for NGC 7212, Contini, Cracco, Ciroi  2012).
An important issue is related with metallicity, which depends on 
mixing with  external matter  enclosed  by merging, new relative abundances of elements ejected by  starbursts,  
gas  trapping into dust grains and   matter from regions close to the AGN accompanying the outward wind
 (Torrey et al 2012). Sub-solar and over-solar relative abundances are both predicted.

Recently,
  Fabbiano et al (2011) reported the presence of two active massive 
black holes, separated by about 490 light years, in the Seyfert galaxy NGC~ 3393. 
They claim that   the  observation  of two black holes very close and  located deep into the bulge
of a regular spiral galaxy, is important to understand AGNs in general and merging galaxies in particular.

NGC~ 3393 was observed by Fabbiano et al (2011)  with the Chandra X-ray Observatory 
camera ACIS-S on 28 February 2004 (ObsID 4868 for 29.7 kiloseconds) and 12 March 2011 
(ObsID 12290 for 70 kiloseconds).  
 Two obscured AGNs appear
in the central regions of NGC~  3393. 
They are most probably  powered by mass accretion.
The lower limits  of the black hole mass are $\sim$8.  10$^5$ \msol
and $\sim$10$^6$ \msol for the NE and SW sources, respectively (Fabbiano et al. 2011).
Simulation results show that the massive black hole separation and the  appearance  of the spiral arms  
created by  the merger,  could resemble those  
of NGC 3393 after about five billion years from collision (Mayer et al. 2007). 

The new discovery turned NGC 3393   into an interesting target to test  merging phenomena.

    NGC 3393 is a  close (z = 0.0125), bright (m$_B$ = 13.1; de Vaucouleurs et al. 1991) 
Seyfert 2, classified as an early type barred galaxy,  appearing face-on. 
It is located in the outskirts of the Hydra cluster
(de Vaucouleurs et al. 1991), covering  more than an arcminute on the 
sky. Cooke et al (2000)   derived from the  redshift a scale of 180 h$^{-1}$ pc arcsec$^{-1}$ 
(h = H$_0$/100 km s$^{-1}$ Mpc$^{-1}$).  Cid Fernandes et al. (2004) give a linear scale of 242 pc/ arcsec.

The strong emission lines in the IUE spectrum suggested that modelling could 
lead to interesting results,
therefore  Diaz, Prieto,  Wamsteker (1988) presented the integrated fluxes of optical and  UV spectra
 providing a first hint to the physical conditions in the nuclear region. 
 Cooke et al. (2000)  explored the NGC 3393 NLR, on the basis of
  the characteristic HST image. 

IRAS fluxes from NGC3393
(Moshir et al. 1990) yield a total IR luminosity of 10$^{10}$ \Lsol and a dust mass of
$\sim$ 5 10$^5$ \msol. If  the IR flux is due  only to star formation, 
 a  star
formation rate  of  $\sim$ 4 \msol yr$^{-1}$ on kpc scale is predicted (Veilleux 1994), corresponding to
 a  rather low  star formation efficiency.
  Fabbiano et al. (2011)  claim on the basis of lack of young stars  in the central region of the galaxy
  that the double black hole   located deep in the bulge of the NGC 3393 AGN
 results from a minor merger event.

 NGC 3393 is a source of a water maser which is the only  resolvable tracer of warm dense molecular gas in the
inner parsec (Kondratko et al 2008).

 Radio emission  at 13 cm taken by the  Australia Telescope Compact Array, 
shows an outflow from the nucleus, 
  not coincident with spiral arms or a bar (Bransford et al. 1998).
The expanding radio lobes sweep up, shock, 
and accelerate gas into shells (Pedlar et al 1985)  which are  fragmented by 
Richtmyer–Meshkov instability due to the underlying turbulence near the shock front.
 Aligning the VLA radio and HST optical central sources,  Cooke et al. found that 
the [OIII] and radio images
superpose. The coincidence of radio and optical emission features indicates that
shocks are at work.

Long slit spectroscopy (Whittle et al 1988) has confirmed morphological and 
kinematic association between the radio lobes and the line-emission gas in Seyfert galaxies.
Therefore,  the high excitation gas extending  beyond the radio source  observed by Cooke et al. (2000) in NGC 3393
will be interpreted as a record of the galaxy collision.
 Whichever the case, the spectra emitted by the clouds will account for both the photoionizing
flux (from the AGN and from the stars) and shocks.

In this paper, we  would like to  investigate   the role of the AGN and of prominent stars
 and their location in NGC 3393,
and  to determine the range and distribution throughout the galaxy of   shock velocities
and  preshock densities. 
In particular, we  would like to  calculate
the relative abundances to  hydrogen of the heavy elements throughout the galaxy,
focusing on  those identified with metallicity (e.g. oxygen in the present case),
because metallicity is  affected by star formation
 and by
outflows of matter, i.e. by the  interactions between the forming galaxy and the
intergalactic medium (Sommariva et al 2011).

We  will investigate NGC 3393  by modelling the  emitted  spectra.  
Formation of shock waves  is  predicted in colliding systems,
therefore we adopt
 for the calculations  the
code {\sc suma}\footnote{http://wise-obs.tau.ac.il/$\sim$marcel/suma/index.htm}
which simulates the physical conditions
of an emitting gaseous nebula under the coupled effect of photoionization from an 
external source and shocks.

The observations of the line spectrum of NGC 3393 by Diaz et al (1988) and  modelling  results
are described in Sect. 2.
HST imaging and spectra, ground-based optical images, long slit spectra presented by Cooke et al
(2000) are   reported and interpreted in Sect. 3.  The continuum spectral energy distribution (SED) calculated
consistently with the line spectra is compared with the observations in Sect. 4.
Discussion and concluding remarks follow in Sect. 5.

\section{The combined UV-optical-IR line spectrum}

Diaz et al. (1988)
 presented  a rich spectrum of emission lines from  NGC 3393, covering the  1200 and 7000 \AA ~ range. 
The line ratios are characteristic of  a Seyfert type 2.
The  UV spectrum shows relatively strong lines, which will be useful to constrain the  models.
  Diaz et al thought, at that time, that NGC 3393 was an early spiral (Sa) 
that most probably did not  have a quasi-stellar nucleus.

Using the low-resolution mode of IUE and its large aperture (20"$\times$ 10")
Diaz et al (1988) detected a strong, flat UV continuum source 
(F$_{\lambda}$ = 1.7 $\times$ 10$^{-15}$ \erg \AA$^{-1}$) spectrum.
Comparing the IUE  flux of He II 1640 with a ground-based  measure of the
He II 4686  taken through a 4"$\times$ 4" aperture, they deduced  a very low 
 reddening of the emission-line spectrum  (E(B-V) $\leq$ 0.07, 
compared to E(B-V) = 0.06-0.09  in the Galaxy).  Yet,
 NGC 3393 is an  IRAS source, indicating the presence of warm dust (Boisson \& Durret 1986).

The optical spectrum was obtained with the Image Dissector
Scanner (IDS) attached to the Boller and Chivens spectrograph at the Cassegrain focus of 
the European Southern Observatory (ESO), La Silla, 1.5m telescope in 1981.
The UV observations  come from the International Ultraviolet Explorer (IUE)
in low dispersion mode through large spectrograph aperture (SWP20148: 400 min, LWP 2844:
110 min and LWP 7602: 361 min).

The combined UV-optical spectrum presented by Diaz et al  shows many lines that
 will be used to  find out the physical and chemical characteristics of NGC 3393
by  modelling  the line ratios.

\subsection{Modelling the line spectra}

\begin{table*}
\caption{Comparison of calculated with observed line ratios to \Hb=1}
\begin{tabular}{lcccccccc} \hline  \hline
   line          &  obs   & m$_{sd}$   & m$_{pl}$                & m$_{bb}$ & m$_{AV}$&m$_{GB}$&FOS\\ \hline
\ Ly$\alpha$ &  25.7  & 127.  &36.           &30.6 &35.4 &-&- \\
\  NV 1252   &   1.58&  28.9 & 1.29         & 0.04 & 1.43&-&-\\
\ SiIV]+ 1403 &   0.63&  40.  & 0.87         & 0.00 & 1.7&-&- \\
\ NIV 1486    &   0.26& 8.   & 0.98         & 0.02 & 0.5&-&-\\
\ CIV 1550    &   4.8 &   76.8 & 3.39         & 0.13 & 4.&-&-\\
\ HeII 1640   &   2.7 &   1.   & 4.5          & 1.88 & 2.3&-&-\\
\ CIII] 1909  &   1.23&   33.8 & 1.84         & 0.43 & 2.&-&-\\
\ NeIV] 2424  &   1.05&  5.2  & 0.35         & 0.035 & 0.3&-&0.32 \\
\ MgII 2789   &  0.78&  2.38 & 0.5          & 2.27   & 2. &-&0.33  \\
\ [OII] 3727  & 1.9  & 3.3   &1.5           &1. & 1.2&    1.65&2.15\\
\ [NeIII] 3869+& 0.7   & 2.5   &1.7           &1.3 & 1.42&1.13&1.2\\
\ [OIII] 4363 & 0.13  & 1.53 & 0.17          &0.06&0.14&0.10& 0.09\\
\ HeII 4686   & 0.3   & 0.074& 0.67          &0.28 & 0.34& 0.29& 0.25\\
\ [NeIV] 4726 & 0.05 & 0.53  &0.014         &5.e-4 & 0.024 &-& -\\
\ [AIV] 4740 & 0.06 & 0.34  &0.1           &0.003 & 0.033& -&-\\
\ \Hb  &      1.     & 1.    &1.            &1. & 1.& 1.& 1.\\
\ [OIII] 5007+& 16.88 & 16.3  &14.24         &13.5 & 14.&13.6& 13.56\\
\ [FeVII] 6087& 0.19 & 1.    &0.15          &0.011 & 0.1&-& - \\
\ [OI] 6300+  & 0.33 & 0.5   &0.137         &0.33 & 0.3&0.41&0.5\\
\ [NII] 6584+& 6.15   & 2.75  &6.3           &3.4 & 4.&4.63& 4.87\\
\ \Ha       & 3.8     & 5.    &2.9           &2.9 & 3.&3.38& 3.\\
\ [SII] 6717  & 0.9   & 0.11  &0.033         &0.37 & 0.3& 1.89& -\\
\ [SII] 6731 & 1.25   & 0.25  &0.073         &0.8 & 0.65& $\uparrow$&-\\
\ [AIII] 7135 & 0.4   & 0.42  &0.7           &0.42 &0.47&-          &-\\ \hline
\ [NeII] 12.8 &-      & 0.9   &0.076        & 0.04&-&-&-\\
\ [NeV] 14.3  & 42.4$^1$ & 0.3   &0.22          &0.014 &-&-&-\\
\ [NeIII] 15.5& 95.$^1$   & 0.33  &1.2          &1.67 &-&-&-\\ \hline
\ \Hb absolute flux $^2$  & -          & 3.8e-3&4.5           &0.59 &1.24&-&-\\ 
\  \Vs(\kms) & -            & 300.  &100.          &600. &-&-&-\\
\  \n0 (\cm3)& -          .& 1500. &3000.         &300.&-&-&-\\
\  $F^3$ & -             & -     &2.3e12        &  -&-&-&-\\
\  \Ts (K)& -             & -     &-       &8.6e4&-&-&-\\
\  $U$ &        -        &-            &-             &1.&-&-&-\\
\  $D$ (cm)& -           & 1.e16 &4.9e16        &1.e17&-&-&-\\
\  str$^4$ & -             &  -    &  0.            &1.  &-&-&-\\
\  RW   &  -             & 0.886 & 0.003          &0.111& -&-&-\\
\ p$_{[OIII]}$ &-              & 4.8   & 77.4           & 17.8 &-&-&-\\ \hline
\end{tabular}

 $^1$ in 10$^{-21}$ W cm$^{-2}$, from  Wu et al. (2011) for the non hidden BLR Sy2 sample; 
 $^2$ in \erg ; 

 $^3$ in number of photons cm$^{-2}$ s$^{-1}$ eV$^{-1}$ at the Lyman limit; 
 $^4$ str indicates infalling (0) or  outflow (1).
\end{table*}

On the basis of the arguments previously mentioned, namely, 1) collision of matter 
originated at the galaxy encounter   and  collision of NLR matter with  the radio outflow from the nucleus,
2) NGC 3393 contains two black holes in its active centre and 3) stars and starbursts are  generally
observed in mergers, 
we have run a grid of models which account for  the
coupled effect of  shocks and photoionization (the primary flux) in both the cases of radiation 
from an active centre and  radiation from the  stars.
A brief summary of the  calculation code is  given in the following (see also Contini et al 2009).

The input parameters: the  shock velocity \Vs, the atomic preshock density \n0 and
the preshock magnetic field \B0 define the hydrodynamical field. They  are  used in the resolution
of the Rankine-Hugoniot equations  at the shock front and downstream. These equations  are combined into the
compression equation which  leads to the  calculations of the density profile downstream.
We adopt  for all the models \B0=10$^{-4}$ gauss.

The input parameter  that represents the radiation field is the power-law (pl)
flux  from the active  centre $F$  in number of photons cm$^{-2}$ s$^{-1}$ eV$^{-1}$ at the Lyman limit,
if the photoionization source is an active nucleus. The spectral indices are $\alpha_{UV}$=-1.5
and $\alpha_X$=-0.7.
 $F$  is combined with the ionization parameter $U$ by
$U$= ($F$/(n c ($\alpha$ -1)) (($E_H)^{-\alpha +1}$ - ($E_C)^{-\alpha +1}$)
(Contini \& Aldrovandi, 1983), where
$E_H$ is H ionization potential  and $E_C$ is the high energy cutoff,
n the density, $\alpha$ the spectral index, and c the speed of light.

If the radiation flux  is   black body radiation from the stars (bb),
the input parameters are the colour  temperature of the  star \Ts
and  the ionization parameter $U$ 
(in  number of photons per  number  of electrons at the nebula).

In models  accounting for the shock,  another important radiation source is
 the secondary diffuse radiation emitted from the slabs of gas heated by the shocks
at relatively high temperatures.

The geometrical thickness of the emitting nebula $D$,
the dust-to-gas ratio $d/g$, and the  abundances of He, C, N, O, Ne, Mg, Si, S, Cl, A, and Fe relative to H
are also accounted for.
 The distribution of the grain radius  downstream
is determined by sputtering,  beginning with an initial  radius of  1. \mum.

The sets of the model input parameters  which best reproduce the data
are  regarded as {\it results}.
The data contain errors both random and systematic and the  calculation results
depend on  the uncertainties of the atomic coefficients  for 12 elements in all the ionization levels.
Therefore, it is generally accepted that the models would reproduce the data within a factor of two.

The calculations  start at  the edge of  an  emitting cloud which corresponds to the shock front.
 (see Contini et al., 2012 for a detailed description of the calculation).
Our models adopt a plane-parallel geometry.

The gas is compressed and thermalized adiabatically at the shock front,
reaching the maximum temperature in the immediate post-shock region downstream :
 ($T\sim 1.5\times 10^5/(V_{\rm s}/100$ km s$^{-1}$$)^{2})$.
The downstream region  is cut into a maximum of 300 plane-parallel slabs
with different geometrical widths, which are
calculated automatically, in order to  follow smoothly the temperature gradient.

Each  line intensity calculated  in each slab downstream  depends  
on the electron density, the electron temperature, and the fractional abundance of the corresponding ion. 
The density of the gas is calculated by the compression equation.
The fractional abundances of   the  ions  is calculated resolving the
ionization equations accounting for the primary flux (the photoionization), the secondary flux and 
collisional ionization.
Both the  flux  from the active centre  and from the stars  as well as the secondary radiation are  calculated by 
radiation transfer throughout the slabs downstream.

The ionization equations are coupled with the energy equation for the calculation of the temperature
in each slab. 
T decreases downstream following the cooling rate by free-free, free-bound and line emission, which  are
calculated in each slab. The line fluxes and dust reprocessed radiation  are integrated throughout the 
geometrical thickness of the cloud downstream.

We have calculated a large grid of models  changing the input parameters
in a consistent way, namely by considering the effect of each of them on
the different line ratios, until a fine tuning of all the line ratios to the data
is obtained. 
 The calculated spectrum  is finally selected    comparing the calculated  with the observed line ratios
in the optical range and
 constraining the precision of the fit  by   discrepancies which  are set at  $\leq$ 20 \% for the 
strongest lines (e.g. \Ly, [OIII]5007+) 
and $<$ 50\% for the weakest ones.

\subsection{Comparison of models with the observations}

We start by modelling the observations of  Diaz et al because many lines   are given in a large frequency range.
 The  observed spectra are presented in Table 1. In the top of the table the UV-optical line ratios 
to \Hb are shown, in the middle we  report the
 [NeV] 14.3 and [NeIII] 15.6 IR line absolute fluxes observed at Earth by  Wu et al (2011).
Notice, however, that in Table 1  the  calculated  line  ratios are referred to
\Hb =1, while the observed IR spectral lines  are absolute fluxes, so we will constrain the models 
 by the [NeV]/[NeIII] line ratios, to avoid  problems of distance, obscuration, etc. 
In the bottom of Table 1 the models are described by the set of the input parameters.

Before discussing the results, we would like to recall that in 
 our models  the shocks are   present also  in radiation 
dominated clouds.
Moreover, we will account  for the  flux from the stars, even if
 Diaz et al. claim that the UV emission  comes  from the nuclear region only and that 
NGC 3393  spectral 
type  is  late enough  as to yield a  negligible stellar contribution in the UV.

Three grids of models were computed adopting   power-law photoionization + shock (m$_{pl}$), 
 black-body  photoionization + shock (m$_{bb}$)  and shock dominated  (m$_{sd}$) clouds.
The  models which best fit the data, are presented in Table 1.

The method adopted for modelling  proceeds  in two main steps. 

 \subsubsection{The physical parameters}

First, we determine the physical conditions of the gas,   adopting the most plausible relative abundances.

We start by calculating the line ratios of oxygen from different ionization levels, [OIII]5007+/\Hb, [OII]3727+/\Hb 
and [OI]6300+/\Hb (the + indicates that the doublet is summed up),
which depend on the fractional abundance of the ions  and on \Te ~ and \ne, and 
the [OIII] 5007+ /[OIII]4363 line ratio which depends strongly on \ne and \Te.
In fact, various oxygen lines are  generally observed in the optical range, and [OIII] 5007+ lines are the 
strongest in Seyfert 2 galaxy spectra. 
 
We try to reproduce the data by changing the input parameters, in particular,
$F$, \Vs, and \n0, considering that n/\n0 ranges between a minimum of 4 (the adiabatic jump) and $>$ 100,
depending on \Vs and \B0, the temperature of the gas depends on the shock velocity, and the photoionization
flux heats the gas to $\sim$ 2-3 10$^4$K and ionizes the gas.

However, the model cannot be definitive until a  satisfactory fit is found also for HeII/\Hb 
 and for CIV/CIII] in the UV.
Both the He  and \Hb lines depend directly on the radiation flux from the  active center (AC)(primary radiation) 
in radiation dominated models,
 while they depend on  secondary diffuse radiation in  collisionally dominated regimes.

The consistent fit of the O, He, and C line ratios is complicated by the fact that, changing
one of the input parameters affects the cooling rate downstream. Consequently, the stratification
of the ions changes and the lines must be recalculated by a different set of  input parameters.

The nitrogen lines NV 1252, NIV] 1486 in the UV and the doublet [NII] 6548+6584 in the optical range
are also important to constrain the models.
However, we do not know which of the lines of the NIV] multiplet is observed, so we will refer 
to the sum of all the terms as an upper limit. NV/[NII] depends on the reddening correction. 
 In case of relatively large FWHM of the line profiles,  [NII] lines are
blended with  \Ha,  increasing their  uncertainty. 
The models which give an acceptable approximation to the data were run with N/H twice solar.

\subsubsection{The relative abundances}

The second step of our modelling method  consists in completing
the fit  of all the line ratios  by changing the relative abundance  of elements which appear 
in the spectrum  by a single line. Generally, these elements  are not strong coolants,
 namely, the corresponding  lines are not strong enough to affect the cooling rate downstream.

Lines such as  MgII and [FeVII] are the  unique representative of the  respective elements, therefore they are
modelled  on the basis of their relative abundance, after the physical conditions of the emitting gas
have been determined by the  line ratios previously mentioned (Sect. 2.2.1).

The [SII]6717 / [SII] 6731  line ratio  depends particularly on the density 
of the emitting gas. 
The [SII]/\Hb line ratios depend on the geometrical thickness of the emitting cloud,
which is mainly  determined by  [OI]/\Hb. So  the calculated [SII]/\Hb  can 
be adjusted to the observed ones by changing the S/H relative abundance.

\begin{table}
\caption{UV line intensities}
\begin{tabular}{lllllll} \hline  \hline
\ line                 & obs & corr  &m$_{AV}$\\ \hline
\ Ly$\alpha$     1215.7&374&440 &885.  \\
\ N V     1238.8	&18&20  & 35.7\\	
\ N V     1242.8&	9&10    &$\uparrow$\\
\ O I     1303&	$<8$& $<$8 & -\\
\ C II     1335&	$<2$ & $<2$ & -\\
\ Si IV     1397&	$<12$ & $<12$  & 42.5\\
\ O IV]     1402&	$\leq$ 18&$\leq$ 18& $\uparrow$\\	
\ N IV]     1486&	$<3$&$<3$ & 12.5\\
\ C IV     1548.2&	64&64 & 100.\\
\ C IV     1550.8	&36&36 & $\uparrow$\\
\ C IV	\ldots	&100&100 & 100\\
\ ?	1593.0	&       9&9&-\\	   
\ He II     1640.7&38& 37 & 57.5 \\	
\ O III]     1663&	$<7$&$<6$ & -\\
\ N III]     1750&	$<9$&$<8$& -\\
\ Si III]     1892.0& $<1$&	$<1$&- \\
\ C III]     1908.7&  34 &34& 50.\\ \hline
\end{tabular}
\end{table}

\subsection{Results}

\subsubsection{AGN, star and shock dominated clouds}

Model  m$_{pl}$  was calculated considering that  the power-law radiation from the AGN
reaches the shock front. 
In the opposite case, 
  the radiation flux from the AC reaches the cloud edge
opposite to the shock front.
This case has been checked and  has been  dropped because many lines were unfitting,
in particular [OI]/\Hb  $>$1.5.
This result suggests that in NGC 3393 the NLR  gas falls toward the
black hole(s).

The absolute fluxes  of [NeIII] and [NeV] observed by Wu et al   (Table 1)
 can be used to  calculate the distance of the emitting gas
from the galaxy centre r. Combining the observed flux at Earth with the flux
calculated at the nebula for   [NeIII] which is a relatively strong line:
 $F_{obs}$ d$^2$=$F_{calc}$ r$^2$$f$,  where $f$ is the filling factor,  d  the distance of NGC 3393 from Earth
(d=54 Mpc, adopting H$_o$=70 km/Mpc) and $F_{calc}$= [NeIII]/\Hb $\times$ \Hb (absolute flux), 
r  results $\sim$ 23/$f^{1/2}$  pc for clouds reached by the AGN radiation.

Table 1 shows that
 the  most acceptable approximation to the observed spectrum is obtained by model m$_{pl}$.
However,  the [SII]/\Hb line ratios are very low compared to observations and
the HeII/\Hb line ratio too high, indicating that the flux from the AC  required to fit 
the [OIII[/\Hb line ratio
is too high. Reducing the flux intensity, however, it will spoil the fit of the  oxygen lines from
three ionization levels. We  check whether the contributions of a shock dominated model m$_{sd}$,
representing  clouds screened from radiation or too far to be reached, and of a  model m$_{bb}$
accounting for radiation from the stars, could help.

The m$_{bb}$ and the  m$_{sd}$ models were selected from  grids with the same criteria
as those used to select  m$_{pl}$.

The best fitting m$_{bb}$ model refers to the case  representing  matter propagating
outwards from the stars. The opposite case was dropped due to unacceptable [OI]/\Hb (=0.02) 
relative to the other  oxygen line ratios, and  very low [SII]/\Hb.
Stars can  drive large-scale winds  ejecting circumstellar gas. 
The winds and radiation fluxes from the stars  enrich and heat  the neighbourhood  medium,  respectively.

The effective temperature of the stars   which leads to  an acceptable fit    of  the calculated to the observed 
optical spectrum  of NGC 3393 (Table 1) is rather high,
\Ts $\leq$  10$^5$  K.  Recall that Wolf-Rayet
 (WR) stars can reach much hotter temperatures than  main sequence stars. A \Ts $\leq$ 140,000 K WC star
can  yield [NeIII]15.6/[NeII] 12.8 $\sim$ 10.
 Similarly, a 120,000 K WN star can reach [NeIII]/[NeII] = 100 (Rigby \& Rieke 2004).
Unfortunately, we do not have the datum of   the [NII] 12.8 line (Table 1) to constrain the models.
 We will see in the next section that
these WR stars,   ionizing and heating the surrounding gas  which is characterised by relatively high \Vs, 
nicely  explain  both the [OIII]/\Hb
line ratio ($>10$) and the high FWHM  observed in the centre of the NGC 3393 galaxy  (Fig. 2).

  Cid Fernandes et al (2004, table 2)  refer to star formation in Seyfert 2.
They present  model synthesis results which show  the percentage in
flux at 4020 \AA ~ of  the  featureless continuum, young, intermediate and old stellar populations.
They find for NGC 3393 a percentage of 83\%  for  old stars and of  4 \% for  young stars, 
in spite  of Gu et al. (2006) claim  that, in general, the  Seyfert 2
nuclear emission luminosity  is mainly  due to young stars
 and that central starbursts are
found in merger Seyfert galaxies (e.g. NGC 6240).

\subsubsection{Densities and velocities of the emitting gas}

The pre-shock densities  relative to model m$_{bb}$ are similar to those found  throughout
other starburst galaxies (100-800 \cm3, Viegas et al. 1999).
On the other hand,  the  model representing the AGN,  and   that representing shock dominated
clouds, show  pre-shock densities  higher by a factor  $\leq$ 10 than
in the  NLR of AGN (300-1000 \cm3, Contini et al 2004 and references therein) and in LINERs (100- 600 \cm3, 
Contini 1997, table 3).  This yields  downstream densities of $\geq$ 10$^4$-10$^5$ \cm3 after compression, 
depending on \Vs and \B0.
We suggest that the high densities adopted  in models m$_{pl}$ and m$_{sd}$  could be   a record of merging.

The  velocities   throughout the NGC 3393 NLR (Cooke et al. 2000, fig.8) are relatively low ($\sim$ 100 \kms) 
similar to those found in the clouds
in the   "region of quiescent regime" (Tadhunter et al. 2000) in the NLR cone of the  Seyfert galaxy
NGC 7212,  most likely a merger,   and similar to the
shock velocities found close  to the Galactic centre (Contini \& Goldman 2011)
most likely a low luminosity AGN (Contini 2011).
Higher velocities are  found close to the  NGC 3393  NLR centre between -20" and 20"
(Cooke et al. 2000, fig.8).

\subsubsection{Metallicities}

In agreement with previous works on NGC 3393 (Diaz et al., Cooke et al.), we find that Mg/H relative 
abundance should be about one third of
  the solar value in the shock dominated and starburst gas, while in the clouds close to the AGN  Mg/H
is about solar.
Solar relative abundances from Allen (1976) are
: C/H=3.3 10$^{-4}$, N/H=9.1 10$^{-5}$, O/H=6.6 10$^{-4}$, Ne/H=10$^{-4}$,
Mg/H=2.6 10$^{-5}$, Si/H=3.3 10$^{-5}$, S/H=1.6 10$^{-5}$, Cl/H=4. 10$^{-7}$, Fe/H=3.2 10$^{-5}$. 
Generally, Mg is  included  into silicate grains, however at shock velocities higher than $\sim$ 200 \kms 
the grains are destroyed by sputtering.
Table 1 shows that the 
 models which overpredict the MgII/\Hb line ratio are m$_{sd}$ and m$_{bb}$, both calculated by
\Vs $>$ 200 \kms.
 Therefore, the Mg depletion could  be due to mixing with external matter
during the merger process. The same is valid for Si.

The models were calculated adopting N/H  twice  solar and C/H =0.7 solar, while the other relative abundances
were solar. 
However, the [SII]/\Hb line ratios depend on the models which are matter bound.
We generally choose the geometrically thickness of the clouds on the basis of [OI]/\Hb.
The larger the cloud, the higher both the [SII]/\Hb and [OI]/\Hb line ratios, because the first ionization
potential of sulphur (10.36 eV) is much lower than that of oxygen (13.61 eV).

In Table 2 we compare the UV lines calculated by the m$_{AV}$ average model
(Sect. 2.3.4) with the FOS blue observations
presented by
Cooke et al (2000, table 9). The observed line ratios to CIV =100 are all overpredicted by the model
by factors $\geq$ 2.
The carbon lines in Table 1 were well reproduced adopting C/H =0.7 solar.
 Table 2 shows that a solar abundance should be adopted for C/H and that  most probably the
reddening correction adopted by Diaz et al. was slightly misleading.

\subsubsection{The average spectrum}

Table 1 shows that the  shock dominated model, although fitting the [OIII] : [OII] : [OI] ratios and their ratios to
\Hb, as well as the CIV$>$ CIII] requirement, overpredicts most of the UV lines by a large factor.
On the other hand, the  model  m$_{bb}$ selected in particular by the agreement of  He/\Hb  line ratios, underpredicts
the UV lines (except the \Ly/\Hb line ratio),  compensating  the results of model  m$_{sd}$ .
Therefore, the spectra  accounting for the different radiation sources were summed up with relative weights (RW)
(which appear in the row before the last in Table 1).
The average model m$_{AV}$ is shown in Table 1 (col. 6)
and is compared with Diaz et al observation data (col. 2),  the ground based observations (col. 7)
and FOS Red Detector observations  (col. 8) reported by  Cooke et al.
(2000, table 7 and table 8, respectively).
In the last column of Table 1 a fit to the ground-based low-resolution spectra, coadded
over a 8"$\times$ 8" region
centered on the nucleus (Cooke et al 2000), are given for comparison.
FOS spectra are also included in the table.

The high RW of the shock dominated model compensates
the very small absolute flux of the calculated  lines, (that can be calculated by  the \Hb absolute flux).
 The RW were selected by the best fit of m$_{AV}$ to the data.
The criteria are the same as those adopted to select  models m$_{pl}$, m$_{bb}$ and m$_{bb}$
(discrepancies $\leq$ 20 \% for strong lines and $<$ 50 \% for weaker lines),
but considering the total combined UV-optical spectrum.

In the last row of Table 1 the resulting  percentage of the absolute [OIII] 5007+  line
for the three models is given, 4.8 \% for shock dominated clouds, 77.4 \% for the AGN 
photoionizad clouds and 17.8 \% for those illuminated by the stars.
The AGN contribution dominates, however both the shock and star dominated  models
are not negligible because they improve the fit to the data.

\begin{figure*}
\centering
\includegraphics[width=12.5cm]{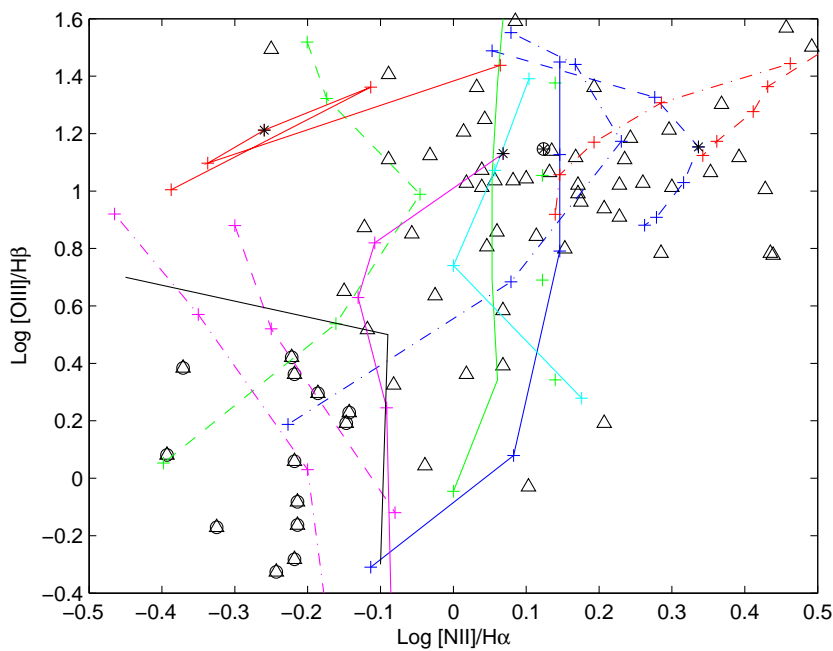}
\caption{Modelling the data from Cooke et al (2000, fig.5).
Black body radiation (+shock) models are represented by magenta lines, 
power-law photoionization (+shock) models are represented by blue-cyan-green lines and
the shock dominated models ($F$=0) by red lines.
The black solid line shows the division between HII regions and AGN, as given in Cooke's  et al.
(2000, fig. 5). The black asterixs represent the models used in Table 1, the encircled asterix
represents the average  model m$_{AV}$. Triangles and encircled triangles reproduce Cooke et al data
for AGN and HII regions, respectively.
Model details are given  in Table 3}
\label{fig1}
\end{figure*}

\section{The physical conditions throughout the galaxy}

\begin{table*}
\caption{The models adopted to explain the data in Fig. 1}
\begin{tabular}{cccccccccc} \hline  \hline
\ \Vs  & \n0  & log $F$ & \Ts    &   $D$         & type & symbol\\
\ (\kms)&(\cm3)& -    & (10$^4$K)&  (cm)         & -     & -     \\ \hline
\  100  & 3000 & 12.36& -        & 2.9e16-3.9e16-4.9e16-5.9e16-7.9e16-1.e17        & pl    & dashed blue \\ 
\ 100   & 3000 & 11.3 & -        & 1.e15-2.6e15-3.e15-4.2e15-1.26e16-2.26e16       & pl &dot-dashed blue \\
\ 100   & 2000 & 9.7  & -        & 6.e13-1.4e14-2.4e14-2.9e15        & pl    & solid cyan \\
\ 300   & 2000 & 11.  & -        & 4.3e14-4.5e14-5.e14-1.1e15-2.5e15        & pl    & solid blue \\
\ 400   & 2000 & 11.  & -        & 5.48e14-5.5e14-5.6e14-5.9e14-6.9e14        & pl    & solid green \\
\ 400   & 2000 & 11.77& -        & 5.8e14-7.e14-1.8e15-4.28e15-1.17e16        & pl    &dashed green\\
\ 600   & 300  & $U$=1.    & 2.6-3.6-4.6-5.6-8.6  &1.e16     & sb  &solid magenta\\ 
\ 100   & 1000 & $U$=0.4    & 3.0-4.0-5.0  &  1.e16   & sb  &dashed magenta\\
\ 100   & 1000 & $U$=0.5    & 2.6 -3.0-4.0-5.0 & 1.e16    & sb  &dot-dashed magenta\\
\ 100-200-300-400-500& 1000& -    &   -      &  $\leq 10^{16}$    & sd   &dot-dashed red\\
\ 100-200-300-400-500& 100& -    &   -       & $\leq 10^{16}$     & sd   & dashed red\\ 
\ 100-200-300-400-500 &   1500& -    &   -       & $\leq 10^{16}$ & sd   & solid red \\ \hline
\end{tabular}
\end{table*}

To obtain some information about the distribution of the physical conditions
throughout the NLR,
we present  in Fig. 1 the  modelling of the spectra reported in the different regions
of NGC 3393   by Cooke et al. (2000, fig. 5).  The lines are the most significant  
([OIII], \Hb, \Ha and [NII]), but they are too few 
 to constrain the models. Nevertheless, we have tried to build up a grid  on the
basis of the models which were used to fit the  spectrum  observed by Diaz et al. (Table 1).
The input parameters of the models are described in Table 3. 
In Fig. 1 we show  the results selected from the grid. They  reproduce the observed [OIII]/\Hb 
and [NII]/\Ha line ratios, corresponding to different physical situations.
The models however, are not constrained by the other line ratios (see Table 1). 
If these lines were observed in the different regions of NGC 3393 NLR, some models would disappear from Fig. 1.
However, Table 3 shows that the input parameters included in the grid, change  smoothly throughout ranges that
are consistent with the  Seyfert 2 NLR physical conditions.  avoiding  unsuitable large jumps.

\subsection{Calculation results}

 Notice that for certain strong line ratios, e.g. [OIII]/\Hb,
a high flux from the AC  and a high shock velocity can give similar results, while the
 other spectral lines change, so the results  are shifted in different positions through Fig 1.  
The same may occur for star  dominated and AGN dominated  models calculated with  different geometrical thickness. 

We adopted N/H twice  solar,  on the basis of the  modelling of the average spectrum.
   Fig. 1 shows that calculating the models  by N/H
 between 1 and 2  solar, we could reproduce all the Cooke et al (2000, fig. 5) data. 

The observations cover the NLR  throughout the galaxy. In this regions the FWHM of the line profiles
show different velocities from $<$ 100 \kms to $\geq$ 400 \kms (Fig. 2). 
Shocks are created  by the
underlying turbulence, which may originate  from collision and merging of galaxies, leading to fragmentation
of matter.  Therefore, the  models are calculated by different geometrical thickness.  

The power-law dominated models are matter bound. 
We have   calculated the line ratios at different distances from
the shock front in each cloud (Table 3), mimicking fragmentation. The results are indicated in Fig. 1 by  
the +  over the curves at the distances  ($D$) reported in Table 3. Lower $D$ correspond to higher [OIII]/\Hb.

High velocity models ($\sim$ 300-400 \kms) with a flux  log($F$)$\leq$ 12 and different
geometrical thickness reproduce most of the cloud spectra (solid blue and green lines).
Low velocity ($\sim$ 100 \kms) power-law flux dominated models (blue dashed and dot-dashed lines)
 reproduce the spectra showing both high [OIII]/\Hb and high [NII]/\Ha line ratios.

Shock dominated  models were calculated for the different \Vs indicated by + over the curves. The values are
given in Table 3. Higher \Vs correspond to lower [OIII]/\Hb.
Characteristic of  shock dominated  models  are the relatively high [NII]/[OIII] line ratios
(red dashed and dot-dashed lines), but at high \n0 $\geq$ 1500 \cm3 (red solid lines)
the [NII]/\Ha line ratios decreases due to the relatively low critical density 
for collisional deexcitation of [NII].

The star radiation dominated models were calculated  at different effective star temperatures (Table 3) 
which are indicated by
+ over the curves.  The higher \Ts the higher [OIII]/\Hb.
The black-body flux   calculated by a high \Ts ($\leq$ 8.6 10$^4$ K) (magenta solid line)
contributes to  reproduce the spectrum observed by Diaz et al (Table 1). At relatively low \Ts
($\leq$ 5 10$^4$ K) the model results follow the
line separating the AGN from the HII regions.
The line ratios from the HII regions (Cooke et al. 2000, fig 5, open circles) are well fitted by the 
star dominated model 
calculated by a black body flux corresponding to \Ts$\leq$ 5.  10$^4$ K.
The  ionization parameter which better fits the  spectrum observed by Diaz et al is $U$=1,
but  lower $U$ (0.4-0.5) characterize the spectra from the HII regions (magenta dot-dashed and dashed lines,
respectively).

\subsection{Interpretation of the data}

\begin{figure*}
\centering
\includegraphics[width=12.5cm]{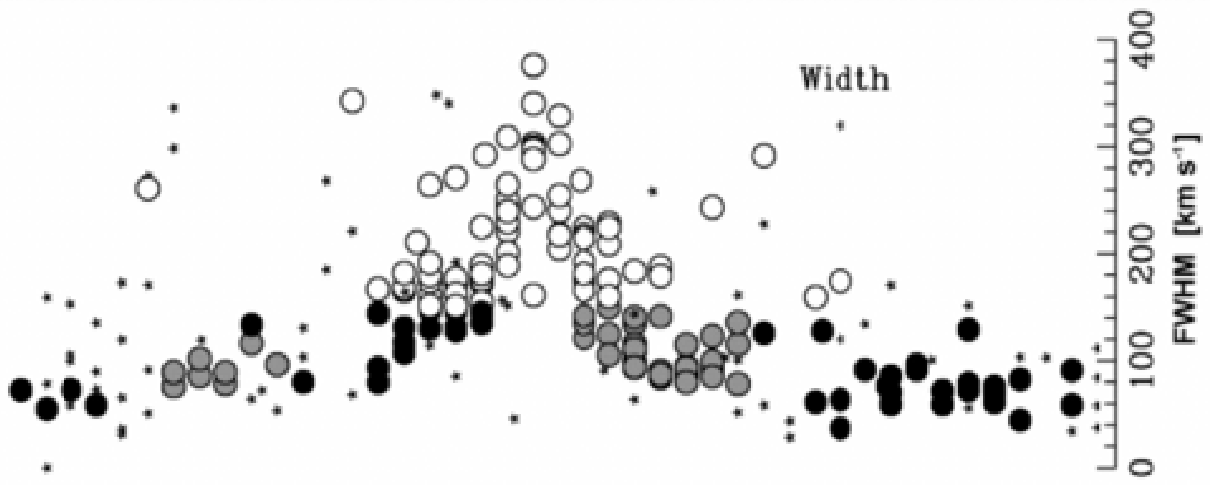}
\includegraphics[width=12.5cm]{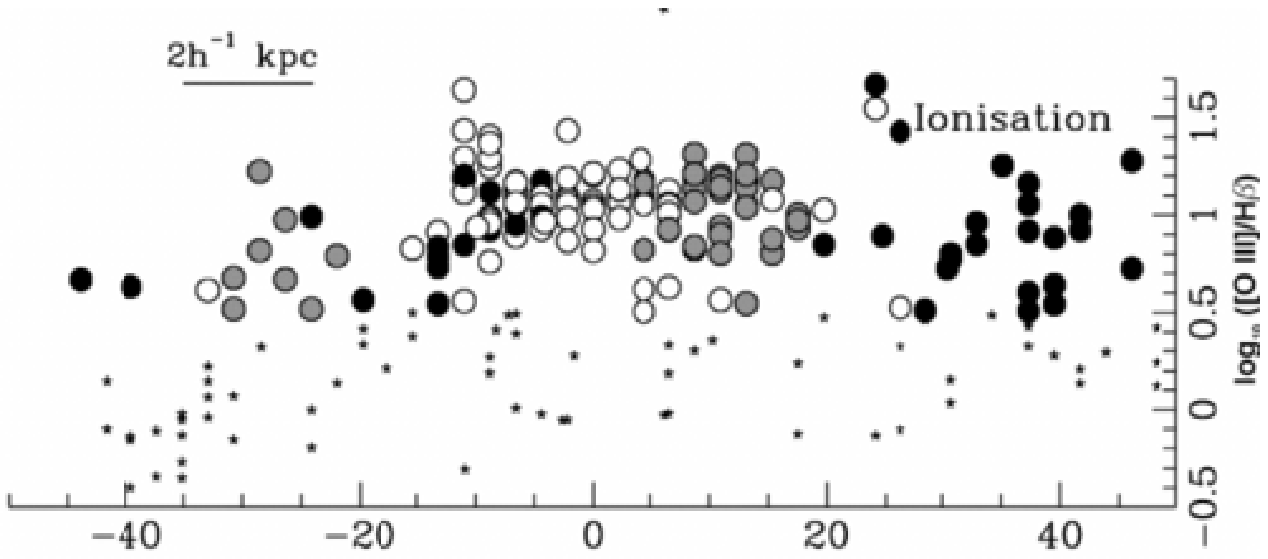}
\caption{Line width (FWHM(\kms)) (top panel) and log([OIII]/\Hb) (bottom panel) as a function of distance from the centre,
adapted from Cooke et al (2000, fig.8).
Circles indicates high ionization gas; open circles  represent locations with the largest line widths; shaded and filled circles
distinguish different spatial regions; small asterisks indicate low ionization gas.
}
\label{fig2}
\end{figure*}

We refer to Fig. 2.  The modelling is  constrained  only by
the FWHM  of the line profiles and the [OIII]/\Hb line ratios. 
We consider that the FWHM  indicate
roughly the shock velocity \Vs in the observed position.

Schematically, we distinguish four types of clouds :
 1) high [OIII]/\Hb, high \Vs ;
 2) high [OIII]/\Hb, intermediate \Vs ;
 3) high [OIII]/\Hb, low \Vs ;
 4) low [OIII]/\Hb,  low \Vs . High [OIII]/\Hb are  close to 10
and the low ones are close to 1. High \Vs are  $\sim$ 400 \kms, intermediate  are   100 $<$ \Vs $<$ 400 \kms,
low \Vs  are $\sim$ 100 \kms.

In  particular,
the velocities as well as the [OIII]/\Hb flux  
peak at the centre of the NLR,  within -0.5 and 0.5 arcsec.
Our models (Fig. 1) show that the emitting gas is ionized by the power-law flux from the
AGN (log($F$) $\sim$ 11- $>$ 12).   Relatively strong shocks  are at work, characterizing the
velocity field (\Vs =400 \kms)
  and yielding fragmentation of the clouds which is revealed by their
 low geometrical thickness  $D$. 
The   emission line fluxes from the clouds illuminated by   stars with \Ts $\sim$ 8.6 10$^{4}$ K  are  blended with 
those  photoionised by the AGN power-law flux. Eventually, also  shock dominated clouds  contribute.

The top  panel of  Fig. 2 shows that  high and low velocities
coexist at slightly larger radii from the centre, while  log ([OIII]/\Hb) are rather 
 constant at two values : $\geq$ 1 and  $\sim$  0. - 0.5 (bottom panel).  The lowest ratios
are  better explained by the  black-body photoionization in  HII regions, 
while the highest ones are  reproduced by both power-law radiation
dominated and shock dominated clouds  corresponding to different \Vs.
Notice that the  AGN dominated model explains the \Vs=100 \kms clouds with high ionization
throughout most of the NLR from East to West.

The FWHM picture is hardly symmetric, even close to the centre, due to clouds
corresponding to higher \Vs on the East. At about -10" the ionization jumps to 
high values (Fig. 2 bottom panel), which can be explained by shock dominated clouds 
with rather low velocities ($\sim$ 100 \kms) and the stars correspond to  the low ionization gas
(small asterisks).
Particularly, at -30" some highly ionised clouds with \Vs$>$ 100 \kms  indicate  shocks outside
the NLR bulk. We interpret them as a collision record.
 On the  West  (Fig. 2),  the shock dominated clouds are less
evident but cannot be excluded.

\section{The continuum SED}

\begin{figure*}
\centering
\includegraphics[width=12.5cm]{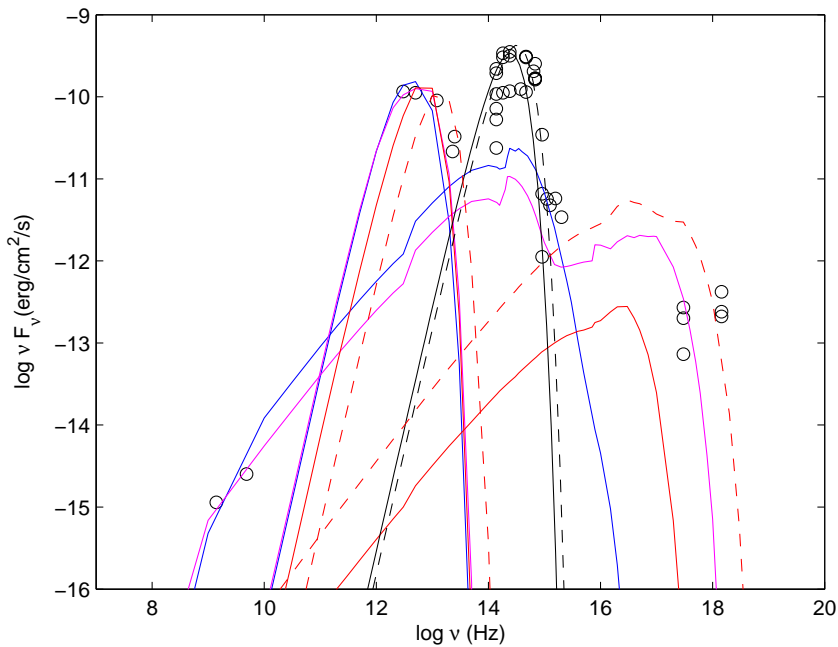}
\caption{Comparison of model results with the continuum SED.
The data from the NED : black circles; blue solid line : model m$_{pl}$;
magenta solid line : m$_{bb}$ ; red  solid line : m$_{sd}$; red dashed  line :
shock dominated  model calculated by \Vs=1000 \kms and \n0 = 1500; black lines : solid:
black body flux calculated by T=3000 K; dashed line : the same for T=4000 K
which represent the old background star population; dot-dashed line : the black body flux
corresponding to 8.6 10$^4$ K;
cyan dashed lines limit  the "obscured window".
}
\label{fig1}
\end{figure*}

The SED of the  NGC 3393 continuum is shown in Fig. 3.
The data come from the NED
 (Levenson et al. 2006; Kinney et al 1993; Munoz Marin et al. 2007 ; Lauberts \& Valentijn 1989;   
 De Vaucouleurs et al 1991; Doyle et al. 2005; Peng et al. 2006; Moshir et al. 1990, Gandhi et al. 2009;
Griffith et al. 1994; Condon et al. 1998). 

We try to reproduce the data by the same models
 as those which lead to the fitting of
the  combined optical-UV spectrum presented by Diaz et al. (1988).
Two lines correspond to each model, one representing  the bremsstrahlung emitted from the gas
and the other showing the dust reprocessed radiation flux.
Recall that the dust reradiation peak shifts at
higher frequencies the higher the shock velocity, because dust grains are heated collisionally by the gas
and radiatively by the power-law  or black-body  flux. 
Collisional heating prevails at relatively high  shock velocities.
Also the  maximum frequency of 
the bremsstrahlung  increases with the shock velocity (Contini, Viegas, Prieto 2004). 

Therefore we  present in Fig. 3 the bremsstrahlung calculated by model m$_{bb}$ which was calculated by 
\Vs=600 \kms. The calculation results give a good fit to the observed soft X-ray data.
Model  m$_{pl}$ calculated by photoionization from the AGN and a relatively low velocity,
\Vs=100 \kms, reproduces the data in the near UV. 

The data in the radio frequency range are few but enough to show that the flux is  a bremsstrahlung
with   some  self-absorption of free-free radiation.  The bremsstrahlung are emitted by the same power-law
 and star black-body dominated  models
which fit the flux in the near-IR and in the soft  X-ray ranges, respectively.
 The radio emission  is optically thick to free-free absorption in luminous infrared galaxies
(Condon et al 1991), in agreement with model calculations.
 Actually, radio synchrotron radiation by the Fermi mechanism at the shock front is not observed in NGC 3393.

Fig. 3 shows that the IR data are well reproduced by the sum of the reprocessed radiation fluxes   calculated
by  m$_{bb}$ and  m$_{pl}$ models which appear in Table 1. 
The IR peak  corresponds to a relatively low dust-to-gas ratio ($\sim$ 10$^{-4}$ by mass).

The peak in the optical-UV range is due to the old background star population.
The data  are nested inside the  black body curves corresponding to  star temperatures  between 3 and 4 10$^3$ K.
 The young hot stars predicted by model calculations have a temperature of 8.6 10$^4$ K. The corresponding
black body flux  peaks in the UV frequency range. However, below the Lyman limit, up to  0.2 keV, 
we have very little observational information; the observations
are difficult because of the heavy absorption by our Galaxy (Fig. 3).

Our results agree with Levenson et al. (2006) who fitted the NGC 3393 nuclear and central spectra
by models which account for gas heated at a  temperature of $\sim$ 3.9 10$^5$ K.
Such a temperature can be found downstream of a shock with \Vs $\geq$ 160 \kms.
 Moreover Levenson et al  report that most of the soft X-ray emission  ($\geq$ 60 \%)
is spread (as well as the optical one) throughout an area of 1700pc $\times$ 770 pc. 
Actually,  the consistent modelling of the line and continuum spectra presented in the previous sections 
indicates that  shocks
are present throughout the whole NLR with velocities between 100 and 600 \kms.

Levenson et al claim that most of the  hard X-rays are  emitted as reflection
continuum emission from the obscured AGN.
Although broad FWHM line profiles were not  observed  in the optical-near IR spectra observed up to now, nevertheless, Fig. 3 
shows that  bremsstrahlung emitted downstream of
shocks corresponding to at least \Vs= 1000 \kms might contribute to the  hard X-rays   emission.
Moreover, the near-IR reradiation by dust calculated by such a high \Vs, would complete the fit of the IR  data in NGC 3393.

\section{Discussion and concluding remarks}

Modelling the spectra in the previous sections we  have obtained the physical and chemical conditions
throughout the  NLR of the NGC 3393 Seyfert galaxy. 
In this paper we searched for   some   evidence of  merging within this galaxy, 
following the results of the observations in the X-ray  presented by Fabbiano et al (2011). They  found
that  two black holes coexist in the centre of the galaxy.

The collision of two galaxies suggests immediately that shock waves will result
leading  in general to  star formation  in the centre 
(Cox et al. 2008) and throughout the galaxy.

Therefore our models account for gas+dust clouds ionized and heated by the
power-law flux from the active centre(s), by  stars and by the shock.

We have found that
 the power-law radiation flux from the  AGN largely
dominates the spectral emission from the gas clouds throughout the galaxy.
The flux ranges between 
$\sim$ 10$^{11}$ and $\sim$ 3 10$^{12}$  
in number of photons cm$^{-2}$ s$^{-1}$ eV$^{-1}$ at the Lyman limit, which
is  characteristic of  the NLR of AGN. Due to their small distance, we treat the
two black holes as the  unique source of  power-law  radiation flux.

The preshock densities are unusually high,  by  a factor of $\sim$ 10  higher than
  in the NLR of Seyfert galaxies. 
Higher densities can reveal that a shock front has crossed
the colliding region  compressing the gas throughout  the galaxy.
The shock wave was most probably the result of different densities on large scales
 in the colliding galaxy gas.

We have searched for the distribution of the shock fronts and their strength
throughout the galaxy and in the  surrounding medium.
By modelling the spectra in the optical-UV range, we have found that
shock velocities range between 100 and 600 \kms. 
The velocities peak in the central region of NGC 3393 (Fig. 2).
 The merging of the  disc matter  
could occur through fragmentation  of dust and gas clouds, typical result of
turbulence at the shock front.
Indeed fragmentation  is predicted by modelling  the emitting cloud spectra
which shows  various  and small geometrical thickness.

Downstream of the shock fronts the gas is compressed and heated to temperatures
which depend on the shock velocity. The spread of shock fronts throughout the NLR
agrees with Levenson et al (2006)  who claim that  NGC 3393 is characterised by 
  extranuclear  soft X-ray emission.

A  galaxy collision  evidence in NGC 7212 was provided by the
observations (Cracco et al 2011) of high [OIII]/\Hb line ratios
and  relatively large [OIII] FWHM profiles  outside  the NLR cone edges.
A similar case  is  found  in NGC 3393.
 Cooke et al. (2000, fig. 8)  show a region to the NE, separated from the
core, which they
identified with the bright patch of [OIII] emission line at $\sim$ 20"E, 20, 20"N
of the nucleus. They claimed it could be {\it some kind of bow shock}.
We suggest that these patches  could be the  record of a collisional event.

Our results show that
stars are present in NGC 3393 with a series
of temperatures, from 3 10$^4$ to 8.6 10$^4$ K. Those  corresponding 
to the highest temperature (\Ts=8.6 10$^4$ K)  are explained by Wolf-Rayet stars.  
The   clouds in the neighbourhood of  the  high \Ts stars
have  pre-shock  densities of 300 \cm3 , shock velocities \Vs  $\sim$ 600 \kms and 
the ionized matter is highly fragmented.  
Fig. 2 shows that they are   most probably located close to the galaxy centre, whereas the oldest stars
accompanied by relatively  low velocities (100 \kms) are located
throughout the NLR.

The results of Cox et al (2008) analysis of the starbursts  that result from
tidal forces in mergers,
indicate that galaxies with  similar mass
 produce the most intense bursts of star formation, while
mergers  of galaxies with  different mass   are not expected to give birth to new stars.
 A starburst  generally appears in the centre of merging Seyfert galaxy, 
between the black holes (e.g. in NGC 6240).
Starbursts  are not predicted in NGC 3393 by the population model analysis  of Cid Fernandes et al (2004)
 who evaluated a 4\%  fraction of young stars  and 83\% of old stars. 
However, we have found that
high velocity clouds illuminated by the  young hot stars contribute
by 17.8\% to  the [OIII] 5007+ line flux compared to low velocity clouds ionised by the AGN which
contribute by $\leq$ 80 \% in the NLR. The   high velocity gas surrounding the  hot stars  provides   
a relatively large fraction of the central  X-ray flux.
The relative fraction of the young hot stars cannot be easily evaluated
 because the  continuum SED of NGC 3393   shows that the black body flux
corresponding to T=8.6 10$^4$ K  peaks in the UV frequency  range where absorption  by our Galaxy is very strong (Fig. 3).

Wolf-Rayet stars of type N  are characteristic of
strong winds spreading nitrogen rich gas in the NLR. In fact, Table 1 indicates that
the emitting nitrogen-rich gas    exhibits the highest
velocities. Our modelling shows that besides  matter ejected from the stars, also the shock dominated 
clouds with \Vs=300 \kms   would better fit the [NII]/\Hb data adopting   N/H 
relative abundance at least twice solar.
These clouds are most probably screened from the star radiation flux by  dusty clumps
 or/and are too far to be reached by the photoionizing flux.

The metallicity of the wind is the same as the  metallicity of gas in the star-forming region from which 
the wind   emerges (Torrey et al 2011). 
Nitrogen rich NGC 3393 results from
the competition between  included low-metallicity gas and enrichment from star formation.
 Yet metallicity is generally related with the oxygen abundance.
In previous sections  we have  found  solar O/H relative abundances  adopting Allen (1976)
 values (O/H= 6.6 10$^{-4}$) in the calculation of the spectra.  The solar abundances
presented by  Anders \& Grevesse (1989) and Feldman (1992) give O/H=8.51 10$^{-4}$, reducing
the metallicity in NGC 3393 to 0.78 solar.
We found also  depletion of Mg by a factor of $\sim$ 3, compared to solar.
Mg cannot be trapped into dust grains which are sputtered throughout the strong shocks.

We suggest that the low O/H and  Mg/H relative abundances show mixing with external matter
as a result of merging. The high N/H is an "indirect" evidence of merging, because  it is a product
of the central stars which, in turn, are created by the collision of near-equal mass galaxies.

In conclusion, by modelling the spectra of NGC 3393 we  have  found some evidence 
of galaxy collision and  merging.
 Our results obtained by consistent calculations are valid, but not definitive due to  scarcity of data.

\section*{Aknowledgements}
I am grateful to the referee for enlightening questions which improved the presentation of the paper.
I  thank Hagai Netzer for  allowing to reproduce
  Cooke et al (2000, fig. 8).
This research has made use of the NASA Astrophysics Data System (ADS) and the NED, which is operated
by the Jet Propulsion Laboratory, California Institute of Technology, under contract with NASA.

\section*{References}

\def\ref{\par\noindent\hangindent 20pt}

\ref Allen C.~W., 1976, Astrophysical Quantities, London: Athlone (3rd edition)

\ref Anders E.,  Grevesse N. 1989, Geochimica et Cosmochimica Acta, 53, 197

\ref Boisson, C, Durret, F.  1986, ESASP, 263, 687

\ref Bransford, M. A.; Appleton, P. N.; Heisler, C. A.; Norris, R. P.; Marston, A. P.
	1998, ApJ, 497, 133

\ref Cid Fernandes, R.; Gu, Q.; Melnick, J.; Terlevich, E.; Terlevich, R.; Kunth, D.; Rodrigues Lacerda, R.; Joguet, B.
2004, MNRAS, 355, 273

\ref Condon, J. J.; Cotton, W. D.; Greisen, E. W.; Yin, Q. F.; Perley, R. A.; Taylor, G. B.; Broderick, J. J.
	1998, AJ, 115,1693

\ref Contini, M., Cracco, V. Ciroi, S. 2012, submitted

\ref Contini, M. 2011, MNRAS, 418, 1935

\ref Contini, M. 2009, MNRAS, 399, 1175

\ref Contini, M. 1997, A\&A, 323, 71

\ref Contini, M.. Aldrovandi, S.M.V., 1983, A\&A, 127, 15

\ref Contini, M., Goldman, I. 2011, MNRAS, 411, 792
 
\ref Contini, M., Viegas, S.M., Prieto, M.A. 2004, MNRAS, 348, 1065

\ref    Cooke, A.J., Baldwin, J.A., Ferland, G.J., Netzer, H., Wilson, A.S. 2000, ApJS, 129, 517

\ref	Cox, T. J.; Dutta, S. N.; Di Matteo, T.; Hernquist, L.; Hopkins, P. F.; Robertson, B.;
 Springel, V.  2006, ApJ, 650, 791

\ref Cox, T.J., Jonsson, P., Somerville, R.S., Primack, J. R., Dekel, A., 2008, MNRAS, 384, 386

\ref      Cox, T.J., Dutta, S. N., Hopkins, P.F.,  Hernquist, L.
Panoramic Views of Galaxy Formation and Evolution
ASP Conference Series, Vol. 399, c 2008 T. Kodama, T. Yamada, and K. Aoki, eds.

\ref  Cracco, V. et al. 2011 MNRAS, 418, 263

\ref     de Vaucouleurs, G., de Vaucouleurs, A., Corwin JR., H.G., Buta, R. J. Paturel, G., and Fouque, P. 1991
    Third Reference Catalogue of Bright Galaxies (New York:Springer)

\ref     Diaz,A.I., Prieto, M.A., Wamsteker, W.
1988, A\&A,195, 53

\ref    Doyle, M. T.; Drinkwater, M. J.; Rohde, D. J.
2005, MNRAS, 361,34

\ref    Duc, P.A., Renaud, F. 2011, ArXiv:1112.1922

\ref    Fabbiano, G., Wang, J., Elvis, M., Risaliti, G. 2011, Natur, 477, 431

\ref Feldman, U. 1992, Physica Scripta, 46, 202

\ref    Griffith, M. R.; Wright, A. E.; Burke, B. F.; Ekers, R. D.	
	1994, ApJS, 90, 179

\ref    Gu, Q.; Melnick, J.; Cid Fernandes, R.; Kunth, D.; Terlevich, E.; 
Terlevich, R.  2006, MNRAS, 366, 480

\ref  Hammer, F., Yang, Y.B., Wang, J.L., Puech, M., Flores, H., Fouquet, S. 2010, ApJ, 725, 542

\ref  	Hopkins, P. F.; Somerville, R. S.; Hernquist, L; Cox, T. J.; Robertson, B; Li, Y. 
	2006, ApJ, 652, 864

\ref Kewley, L. J.; Geller, M. J.; Barton, E. J.  2006, AJ, 131, 2004

\ref    Kinney, A. L.; Bohlin, R. C.; Calzetti, D.; Panagia, N.; Wyse, Rosemary F. G.
	1993, ApJS, 86, 5

\ref     Lauberts, A. and Valentijn, E. A. 1989
The surface photometry catalogue of the ESO-Uppsala Galaxies,  Garching Bei Munchen: European Southern Observatory

\ref    Levenson, N. A.; Heckman, T. M.; Krolik, J. H.; Weaver, K. A.; Zycki, P. T.
	2006, ApJ, 648, 111

\ref    Mayer, L. et al.  2007, Science 316, 1874  

\ref Moshir, M. et al. 1990 in $IRAS$ Faint Source catalogue, version 2.0(JPLD-10015 8/92; Pasadena:
Jet Propulsion Laboratory)

\ref    Mu\~{n}oz Mar\`{i}n, V. M.; Gonz\`{a}lez Delgado, R. M.; Schmitt, H. R.; Cid Fernandes, R.; 
P\`{e}rez, E.; Storchi-Bergmann, T.; Heckman, T.; Leitherer, C.
	2007, AJ, 134, 648

\ref Pedlar, A.; Unger, S. W.; Dyson, J. E. 1985, MNRAS, 214, 463

\ref    	Peng, Z.; Gu, Q.; Melnick, J.; Zhao, Y.
2006, A\&A, 453, 863

\ref Rigby, J.R., Rieke, G.H. 2004, ApJ, 606, 237

\ref    Shir, M. et al. 1990,
Infrared Astronomical Satellite Catalogs, The Faint Source Catalogue, version 2.0
1990 

\ref    Sommariva, V. et al. 2012, ArXiv:1112.2403

\ref Springel, V.; Di Matteo, T.; Hernquist, L.	2005, MNRAS, 361, 776

\ref   Tadhunter, C. N.; Villar-Martin, M.; Morganti, R.; Bland-Hawthorn, J.; Axon, D.	
	 2000, MNRAS, 314, 849

\ref  Torrey, P., Cox, T. J., Kewley, L., Hernquist, L. 2012, ApJ, 746, 108  

\ref    Two  Micron all Sky Survey Team
    2Mass Extended Objects. 2003 final release

\ref Veilleux, S.; Cecil, G.; Bland-Hawthorn, J.; Tully, R. B.; Filippenko, A. V.; Sargent, W. L. W.
	1994, ApJ, 433, 48

\ref Viegas, S.M., Contini, M., Contini, T. 1999, A\&A, 347, 112

\ref Whittle et al. 1988 ApJ, 326, 125

\ref Wu, Y.-Z.,  Zhao, Y.-H.,  Meng, X.-M. 2011, ApJS, 195, 17


\end{document}